**D. V. DOBRYCHEVA**[1], Senior Researcher, PhD in Phys. & Math.

**I. V. KULYK**[1], Senior Researcher, PhD in Phys. & Math.

**D. R. KARAKUTS**[1,2], Junior Researcher, PhD student

**M. Yu. VASYLENKO**[1,2], Junior Researcher, PhD student

**Ya. V. PAVLENKO**[1], Doctor of Sciences

**O. S. SHUBINA**[1,3], Researcher, PhD in Phys. & Math.

**I. V. LUK'YANYK**[4], Head of Department, PhD in Phys. & Math.

[1]Main Astronomical Observatory of the National Academy of Sciences of Ukraine

27 Akademika Zabolotnoho Str., Kyiv, 03143 Ukraine

[2]Institute of Physics of the National Academy of Sciences of Ukraine

46 avenue Nauka, Kyiv, 03028 Ukraine

[3]Astronomical Institute of Slovak Academy of Sciences

Tatranská Lomnica, 059 60 Vysoké Tatry, Slovak Republic

[4]Astronomical Observatory of Taras Shevchenko National University of Kyiv

3 Observatorna Str., Kyiv, 04053 Ukraine

# VISUAL INSPECTION OF POTENTIAL EXOCOMET TRANSITS IDENTIFIED THROUGH MACHINE LEARNING AND STATISTICAL METHODS

*In this work, we explore several ways to detect possible exocomet transits in the TESS (The Transiting Exoplanet Survey Satellite) light curves. The first one has been presented in our previous work, a machine learning approach based on the Random Forest algorithm. It was trained on asymmetric transit profiles calculated as a result of the modelling of a comet transit, and then applied to real star light curves from Sector 1 of TESS. This allowed us to detect 32 candidates with weak and non-periodic brightness dips that may correspond to comet-like events. The aim of this work is to analyse the events identified by the visual inspection to make sure that the features detected were not caused by instrumental effects. The second approach to detect possible exocomet transits, which is proposed, is an independent statistical method to test the results of the machine learning algorithm and to look for asymmetric minima directly in the light curves. This approach was applied to β Pictoris light curves using TESS data from Sectors 5, 6, 32, and 33. The algorithm reproduced nearly all previously known events deeper than 0.03 % of the star flux, showing that it is efficient to detect shallow and irregular flux changes in the different sectors of the TESS data and at the different levels of noise.*

*The combination of machine learning, visual inspection, and statistical analysis facilitates the identification of faint and short-lived asymmetric transits in photometric data. Although the number of confirmed exocomet transits is still small, the growing amount of observations points to their likely presence in many young planetary systems.*

## 1. INTRODUCTION

The infrared and millimeter-wavelength surveys conducted in recent years show that around 20 % of nearby old stars (several Gyr) possess debris disks, which are thought to be analogues of the Kuiper belt of our Solar system ([26] and references herein). For younger stars, it is assumed that an even higher percentage of them are surrounded by less evolved dusty debris disks [20]. The parent star's radiation pressure and collisions induce a short lifetime for the observable dust in disks, necessitating constant replenishment. Recent direct JWST observations of the β Pictoris disk point out that ongoing comet activity, along with the collision events, could be the mechanism feeding the disk with dust [24]. In addition, Gaia DR3-based kinematic analysis of nearby stellar encounters suggests that close flybys may perturb the outer cometary reservoir of planetary systems, including β Pictoris, increasing the inward flux of comets and planetesimals [5].

The existence of an extrasolar comet population was revealed through observations of anomalous absorption features in the spectra of approximately 20 young A-stars [23]. The short-term variability in the shape and depth of the absorption line, often redshifted and superimposed on the H and K CaII photospheric lines, indicates the presence of bodies surrounded by mini-atmospheres, moving at close distances from the star and randomly crossing the line of sight [9, 10]. This phenomenon also causes small changes in the brightness of a parent star, as predicted and theoretically calculated by [14]. More recently, [6] analysed over 9000 HARPS spectra of β Pictoris and detected only two Na I events coinciding with strong Ca II lines, which shows that sodium appears only in the most active exocomets. They also found year-long Ca II absorption and rapidly accelerating events likely linked to comet destruction near periastron, revealing a broad range of exocomet behavior within the system. According to the recent review by [1], after a protoplanetary disk turns into a debris disk, the star may still experience a slow but continuous fall of small ice bodies from its outer reservoirs. Most of the known cases are related to A-type stars, yet there is no reason to exclude such processes around stars of other spectral types.

The photometric manifestation of exocomet transits was found later when high-quality photometric time-series data were collected by the Kepler and TESS (The Transiting Exoplanet Survey Satellite) space missions [18, 25]. The aperiodic transits discovered in the light curves of several stars have a pronounced asymmetric shape with a steeper drop in the starlight at the beginning of the transit event and a slower increase in the brightness of the star at the transit egress [8, 12, 13, 19, 22, 30]. The asymmetry of the exocomet profile shape is consistent with the theoretical predictions of a comet passage across a star disk [14].

Recently, [8] proposed an automatic algorithm designed to search for the asymmetric transits in light curves, and the method was applied to the Kepler telescope database. After analysing about 68,000 potential transits, the authors found only two events, indicating possible cometary transits, in addition to the few ones that had already been found previously. The automatic algorithm was utilized to search for exocomet transits in sectors 1—26 of the TESS database, and five new transits were identified [17]. The authors estimated the occurrence rate of exocomet transits at a level of $2.64 \times 10^{-4}$ stars$^{-1}$years$^{-1}$, but the depths of such transits are shallow. For most of the transits identified, they are less than 0.1—1 % of the star's flux.

In this work, we analyse light curves through visual inspection of the objects previously identified by our machine learning methods as potential candidates with asymmetric transits, and we further strengthen the verification by applying statistical methods to search for asymmetric minima in the TESS database. We are looking for sporadic aperiodic transit events, which can also be caused by passing small bodies like exoasteroids or long-period planets in front of the star. The specific statistical features characterizing the morphology of the transit profiles are used as a discriminator to separate the transit events attributed to a comet-like body. For this reason, we use the light curves gathered by the TESS space mission and stored in the Mikulski archive for Space Telescopes. Unfortunately, only a few systems have exhibited exocomet activity compared to the more than 5000 confirmed exoplanet systems at this time. This limited number of detections is insufficient for machine learning methods to train a model to search for exocometary activity effectively. We need large datasets to allow the machine learning algorithms to recognize patterns and make accurate predictions. The current scarcity of confirmed exocometary events poses a challenge for developing robust models in this area. Expanding the number of known systems with detected exocomet transits would be essential for improving the training and performance of such methods.

## 2. DATA COLLECTION AND THE MACHINE LEARNING ANALYSIS METHOD

For our analysis, we employed the TESS 2-minute short-cadence Presearch Data Conditioning (PDC SAP) light curves from sector 1 generated by the science analysis pipeline of the Science Processing Operation Center (SPOC, [7, 15]). The PDC segment of the SPOC applies a series of corrections to the light curves taking into account the instrumental artifacts caused by the focus or pointing instability, discontinuities in the data due to radiation events in the CCD detectors and so on; removing isolated outliers and correcting fluxes for aperture effects, such as field crowding or the fractional loss of target flux due to star centroid drifting [7].

The precision of each light curve is estimated with the Combined Differential Photometric Precision (CDPP) metric, developed by the Kepler space telescope team and later applied to TESS data, which is the signal-to-noise ratio after smoothing and whitening light curves and measured in parts per million (ppm) [4, 27]. For instance, a light curve with a CDPP of 20 ppm and a 3-hour transit duration can detect a transit of 20 ppm depth and corresponding duration span at a 1 sigma level. This metric is convenient for evaluating light curve precision throughout data processing.

In order to find and identify the asymmetric decrease in the star brightness caused likely caused by a comet-like body , we apply the classical ML Random Forest method [3, 28], for which we constructed the classifier using the Python module Scikit-learn [21]. Since we have extensive experience applying machine learning methods to extragalactic astronomy [11, 29], as well as to objects with extremely small training samples [2], we also examined some other methods, such as Logistic Regression, Support Vector Machines, k-Nearest Neighbors, but the Ran-dom Forest initially showed better performance, prompting us to focus on it. Analysing why the Random Forest method showed better results, we came to the conclusion that it is an ensemble approach that combines the results of multiple decision trees. Each tree can capture complex non-linear relationships between features, which may arise due to asymmetric brightness variations, noise, or random fluctuations in light curves.

To prepare a training sample for the machine learning method, we simulate the artificial asymmetric transit events since only a few exocomet transits have been confirmed so far in the time series data [28]. In order to retrieve the shapes of artificial exocomet transits, we calculate the attenuation of the star flux when the modeled comet 'moves' across the star disk. The model of the comet dusty coma was created using the Monte-Carlo method, incorporating known physical properties of some solar system comets. More details explaining the process of transit simulation can be found in [16], where we also clarify our methodology and detailed descriptions of the required parameters.

We encapsulated artificial transit profiles into the light curves from sector 1 TESS data randomly to produce a simulated training dataset. Figure 1 demonstrates the procedure of the transit encapsulation in detail. The top panels show the artificial transit profile and the light curve, both randomly selected from the corresponding datasets. The bottom panels depict the segment of the light curve with the transit encapsulated and a zoomed span of the light curve in the transit vicinity. You can read more about the application of machine learning methods in our work [3]. Here we present only a brief description. To predict possible transits with the asymmetric shapes in the light curves, we use a binary classifier that divides the light curves into two classes: 'exocomet candidate' and 'non-candidate'. Eventually, the training sample contains 20,000 light curves: 10,000 are labeled '0', marking 'non-candidate', and 10,000 are labeled '1', marking 'exocomet candidate'.

Taking into account that the observed exocomet transits produce a depth of less than 1 % of the star's brightness, we limited the sample by selecting light curves with a predetermined precision. We created two target samples by selecting the light curves with CDPP less than 40 ppm and 150 ppm, respectively. This resulted in about 2000 light curves with CDPP under 40 ppm and 9000 with CDPP under 150 ppm. Each sample is accompanied by a corresponding training set. We used the TSFresh Python package for feature extraction, which helped in refining the Random Forest model. The training set was used to train the model, resulting in a high accuracy rate of about 96 %, with precision, recall, and an F1-score all above 95 %, effectively separating potential exocomet candidates from non-candidates. 12 possible candidates were identified from the target set containing light curves with a precision of less than 40 ppm CDPP. Despite a higher rate of unclassifiable curves in the target data set with a precision limit of 150 ppm CDPP, our model still pinpointed 20 potential candidates.

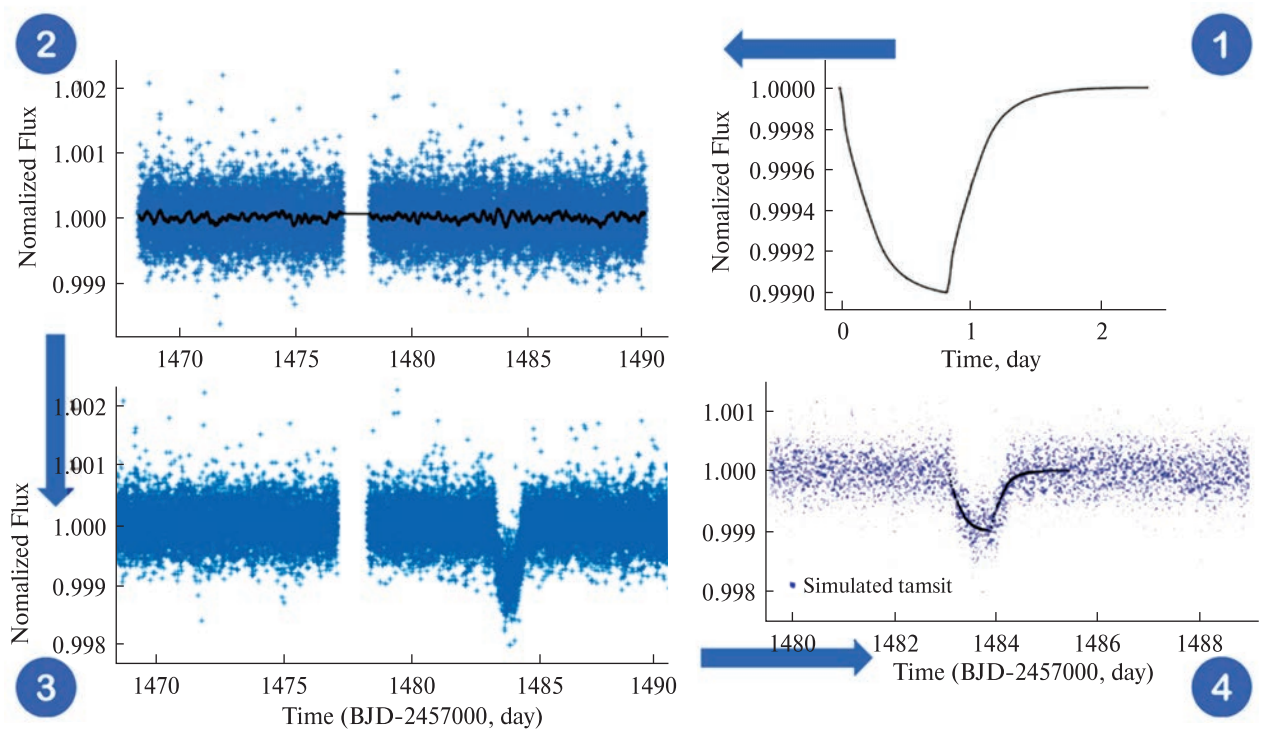

***Figure 1***. Step-by-step procedure of the transit encapsulation in the TIC 020209388 light curve. The X-axis and Y-axis specify time in days and normalized flux, respectively. Top panel: *1* — simulated transit profile, *2* — PDC_SAP normalized flux of the star from sector 1 TESS dataset, both selected randomly; bottom panel, *3* — the light curve of the star with the encapsulated transit, *4* — zoomed span of the light curve in the vicinity of the encapsulated transit

All 32 identified candidates were manually verified through a visual inspection of their light curves to confirm a potential exocomet transit.

## 3. VISUAL INSPECTION OF THE IDENTIFIED CANDIDATES AFTER APPLYING THE MACHINE LEARNING APPROACH

We conducted a visual inspection of all 32 candidate exocomets. It is important to mention that the artifact (see Figures 2...9), which occurs in all light curves of sector 1 TESS data in the time span between 1347 and 1349 BTJD (Barycentric TESS Julian Date), has been taken into account when working with feature extractor for machine learning method Random Forest.

Among the identified candidates, several light curves exhibit periodic and symmetric dips in brightness that are typical of exoplanet transits rather than exocomet events. In particular, three stars show repeating minima with periods shorter than 27 days — the length of TESS sector — and the depth and symmetry of these features are fully consistent with planetary transits.

Figure 2 shows the transit event in the TIC 214517366 light curve. The interval in the transit vi-cinity is highlighted with green lines and zoomed out in Figure 3 for clarity. The zoomed figure shows the considerable depth of the event, up to 4 %, and the symmetric profile. However, the absence of periodicity in the light curve suggests that the object may be a planet with an orbital period exceeding the observation window. This deep event can also be attributed to the occultation of the star by the solar system asteroid. We have found four such single transits among the 32 candidates selected. Figure 4 shows the TIC 272322827 light curve with quasi-periodic star variations. Machine learning methods identify this light curve as one that has the potential exocomet transit. Visual inspection suggests that the feature highlighted with green lines is most likely identified as an asymmetric transit; however, it is related to intrinsic stellar activity and the light curve artifact in sector 1. The artifact interrupts the sine wave, creating a steep drop in the signal and an asymmetric feature around the time moment of 1350 d, leading to false identification. We have found seven such examples among the candidates for exocomets.

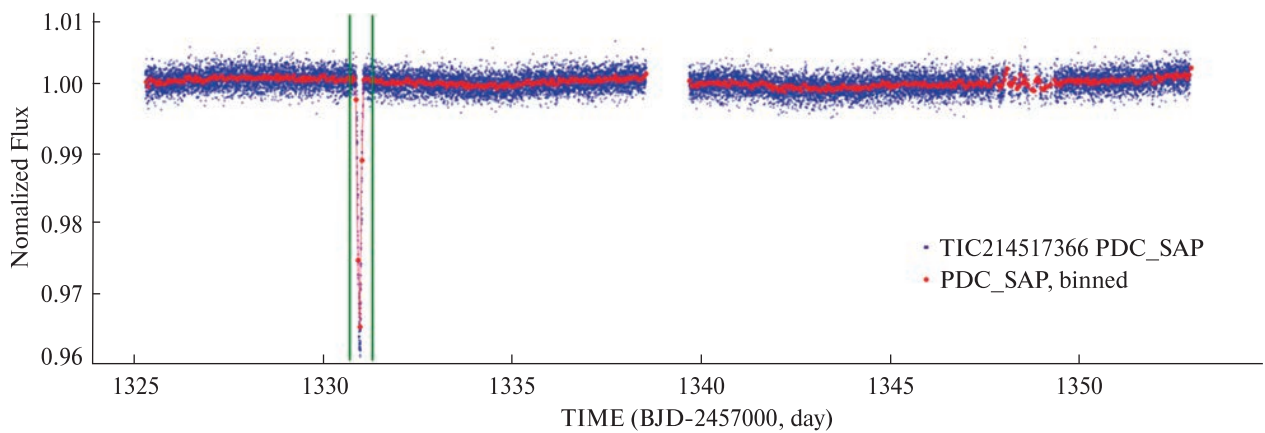

**Figure 2**. The TIC 214517366 light curve demonstrates a deep transit that may be due to occultation by a long-period planet

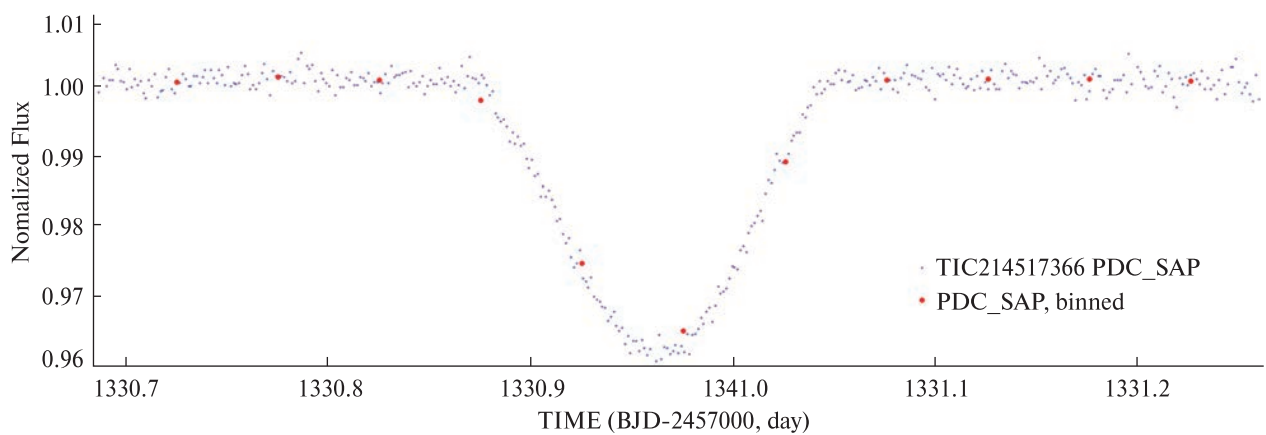

**Figure 3**. Segment of the TIC 214517366 light curve in the transit vicinity

Figure 5 shows a case of the false identification in the TIC 215053453 light curve, which the machine learning algorithm flagged as potentially containing an event near the time moment of 1325 d. Closer inspection suggests that this feature is most likely due to noise, arising from instrumental factors at the sector edge, rather than a planet or a comet transit. The rest of the light curve shows minimal variation, further supporting the conclusion that this is not an astrophysical event. Similar cases occur in other light curves, where short, non-repeating flux variations appear at the beginning or end of the observation window. These artifacts can mimic asymmetric dips and lead to false detections of exocomet-like events. In total, fifteen such cases were found among the selected candidates. And only two interesting events were identified as potential exocomets. The first can be seen in Figure 6. It shows the light curve of TIC 441042358 with the two dips marked with green lines and zoomed in Figure 7. The deeper one, with a depth of up to 0.2 %, is characterized by a slightly asymmetric shape at the transit egress. The mild asymmetry and absence of periodicity suggest that the event can be caused by an exocomet. Additionally, two close dips can be attributed to successive transits of two planets or comets. Further monitoring is vital to examine these anomalous transit events. Unfortunately, only one light curve for TIC 441042358 has been available in the TESS database so far. The second case of a suspicious transit event can be seen in Figure 8, which presents the TIC 121490734 light curve, the zoomed part of it around the time moment of 1342.2 d you can see in Figure 9. As with the event mentioned above, additional observations of the star are required to search for other similar events, which can confirm the presence of exocomet activity.

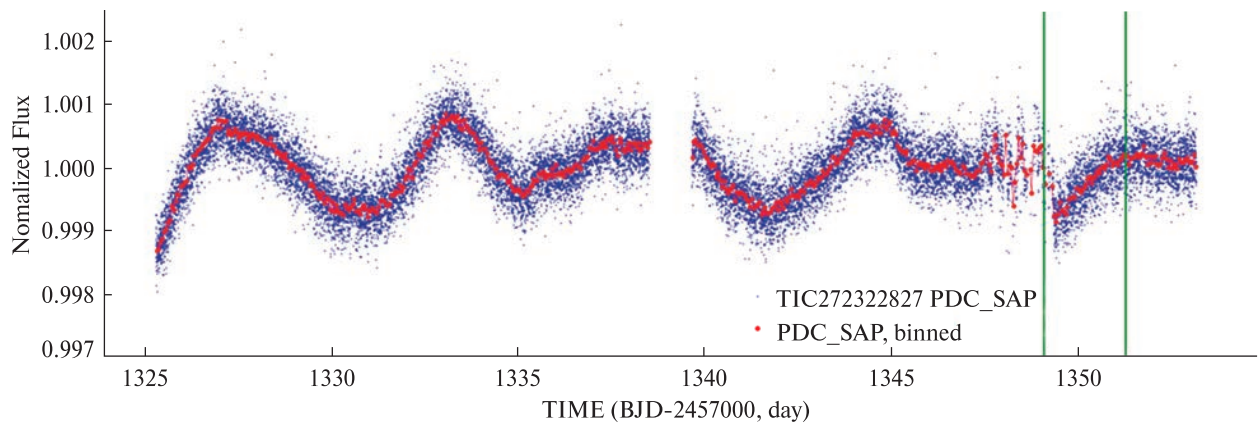

***Figure* 4**. The light curve of TIC 272322827 demonstrates the star's own oscillations and asymmetric feature marked with green lines, which is likely caused by the artifact

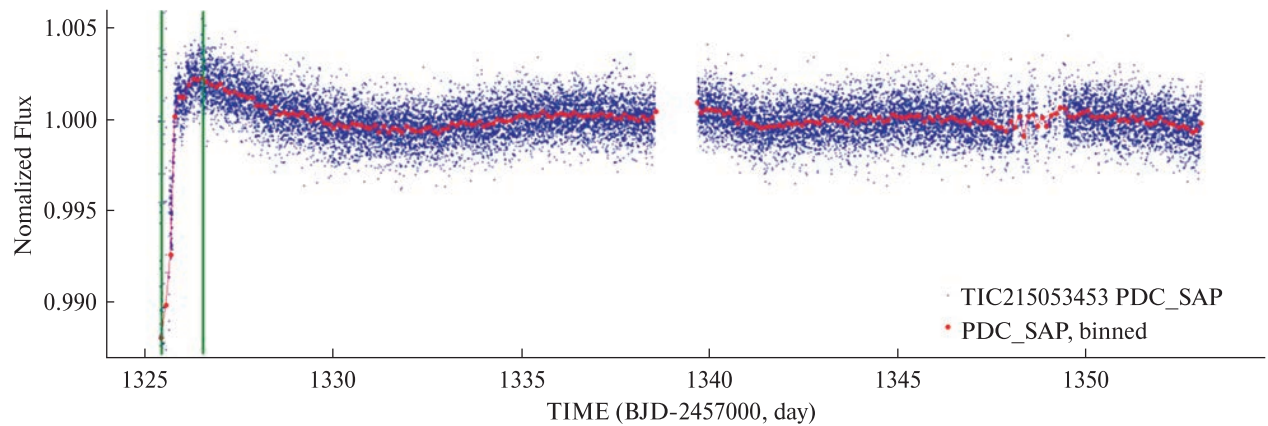

***Figure* 5**. The light curve of TIC 215053453 demonstrates noise near the sector edge, marked with green lines, likely due to instrumental or observational factors, which was flagged as a comet transit

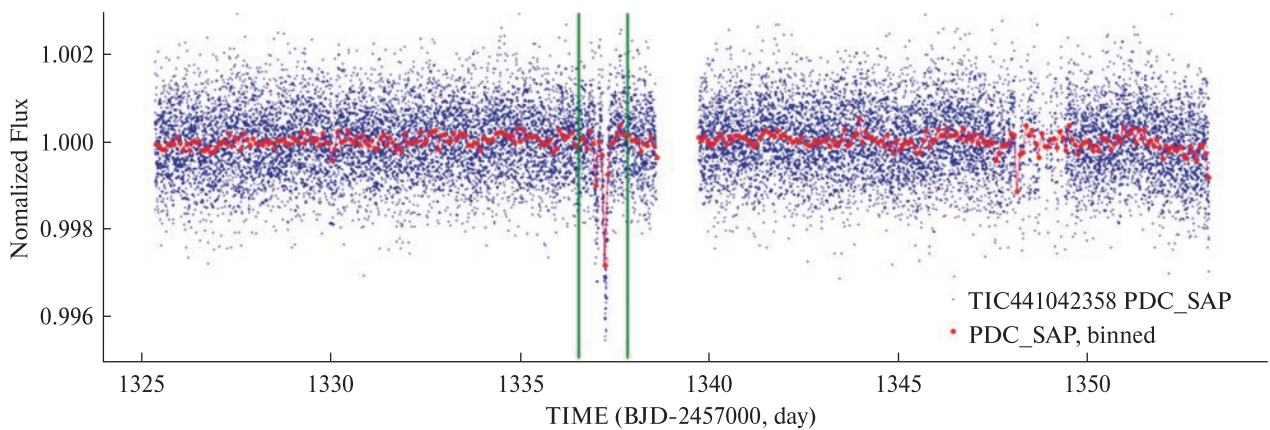

***Figure* 6**. The TIC 441042358 light curve demonstrates an event flagged as a potential exocomet transit marked with green lines

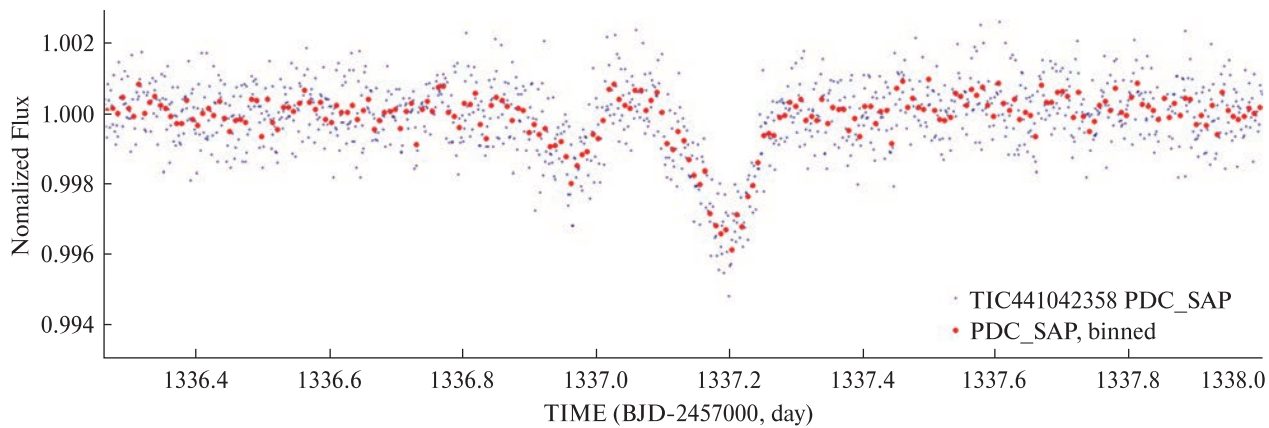

***Figure* 7**. Zoomed-in section of the TIC 441042358 light curve demonstrates an anomalous transit, which can be caused by successive transits of two planets or comets

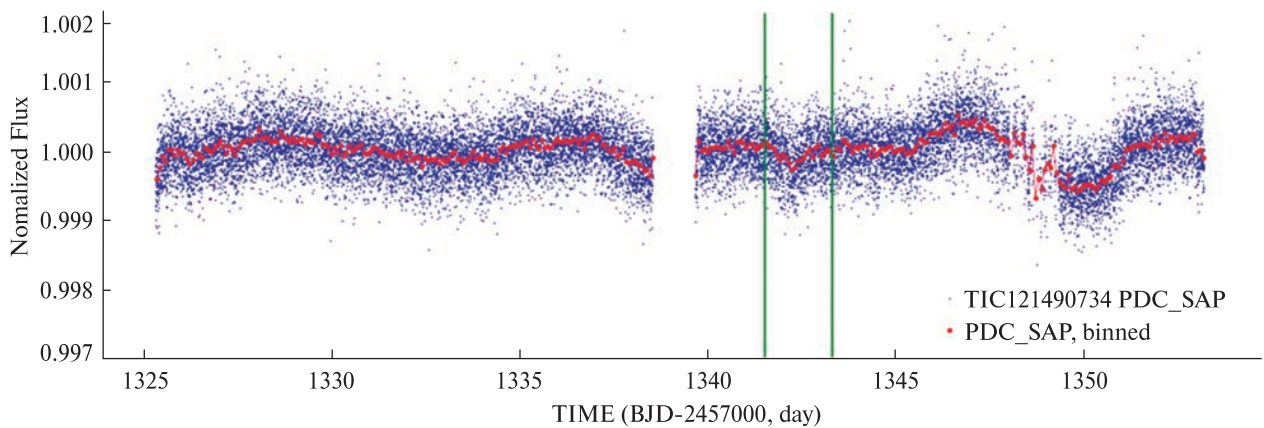

***Figure* 8**. The TIC 121490734 light curve demonstrates a flagged event as a potential exocomet transit

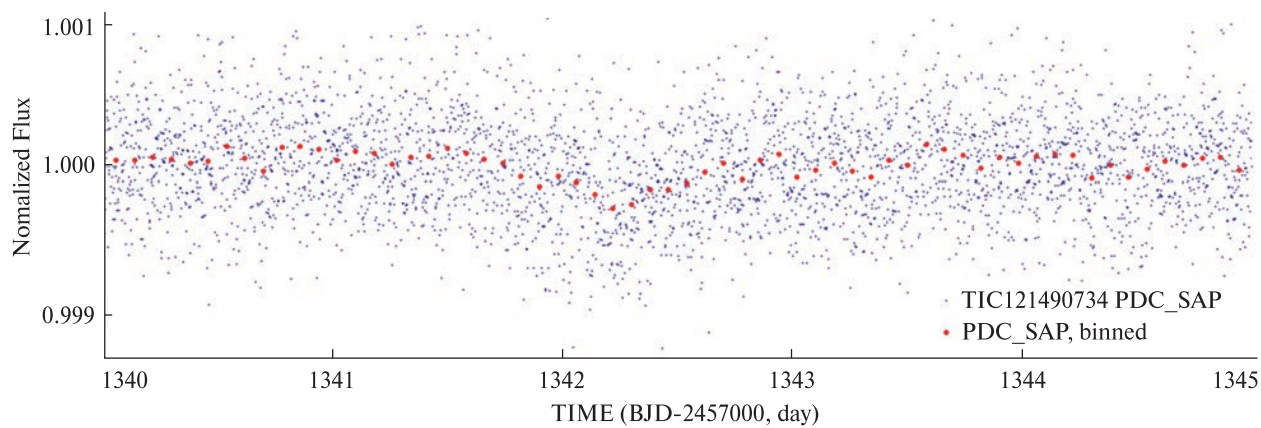

***Figure* 9**. Zoomed-in section of TIC 121490734 light curve demonstrates a potential exocomet transit

## 4. STATISTICAL APPROACH

To reinforce the results of machine learning methods, we decided to use an additional statistical method. In this approach, we load the raw TESS data in FITS format and perform a normalization of the stellar flux. After normalization, the long-term variations were removed with the Wotan algorithm, which produced flattened light curves. To make the transit features more visible, we additionally smoothed the data to suppress small-scale noise and preserve the overall shape of possible exocomet events. To achieve this, we have decided to use a centered rolling window to calculate the mean symmetrically around each point, ensuring that the data hasn't temporally shifted. The method was tested on the β Pictoris light curves from Sector 5, which contain known exocomet events from the previous investigations [13, 19, 30].

Empirically, the window size *w* of 90 points was found to provide good results, though it can be further optimized, depending on the specific star's pa-

rameters. The centered moving average at time *t* is given by:

$$\overline{f}(t) = \frac{1}{w} \sum_{k=-w/2}^{w/2} f(t+k) \,,$$

where  is the centered moving average trend flux at time *t*, *f(t)* is the trend flux value at time *t*, *w* is the window size, representing the number of points included in the moving average calculation, *k* is an index variable that runs from $-w/2$ to $+w/2$, indicating the range of data points included in the average, centered around the current point *t*.

To identify time-series intervals that may correspond to transit events, we calculated the differences between neighboring points of the binned data and searched for segments where a substantial negative trend is followed by a positive one. The differenced data of moving averages for trend flux is given by:

$$y(t) = \overline{f}(t) - \overline{f}(t-1) \,,$$

where *y(t)* is differenced data, $\overline{f}(t)$ is the centered moving average trend flux at time *t*, $\overline{f}(t-1)$ is the centered moving average trend flux for lagged time $t-1$.

An example of a moving average trend with a lag of 1 point for β Pictoris exocomet transit is shown in Figure 10. Negative trend is given in red and positive — in green. The transit minimum is reflected with a dotted blue line. Moreover, the flux of the asymmetric transit itself is presented below, where red dots show the normalized flattened flux, the green line represents the moving average flux of the transit event, while the ingress, minimum, and egress points are blue markers.

This way, flattened light curve was transformed into moving trends for subsequent analysis, specifically to identify potential exocomet transiting events. A key challenge is the variable width of these transit events, which necessitates searching a range of left and right wing combinations. To significantly reduce the computational load, the search is focused only on points where the trend's sign changes from negative to positive.

For the points where the trend changes its sign, we calculate the maximum width of the negative segment of the differenced moving-average flux to the left, allowing up to 5 % positive fluctuations (i.e., 95 % of points remain negative). We focus on events

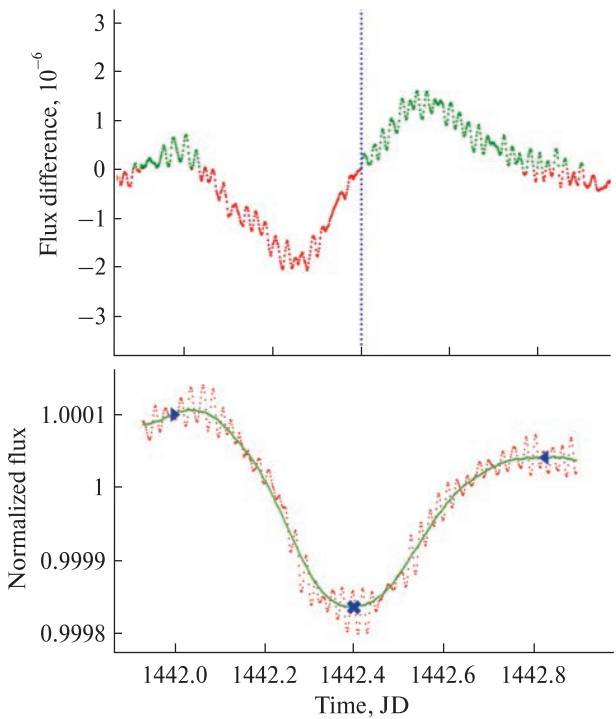

**Figure 10.** Top plot — differenced data of moving averages for trend flux. The bottom plot — normalized flattened flux (red), moving average flux (green), and points of interest: transit's ingress (>), minimum (X), and egress (<)

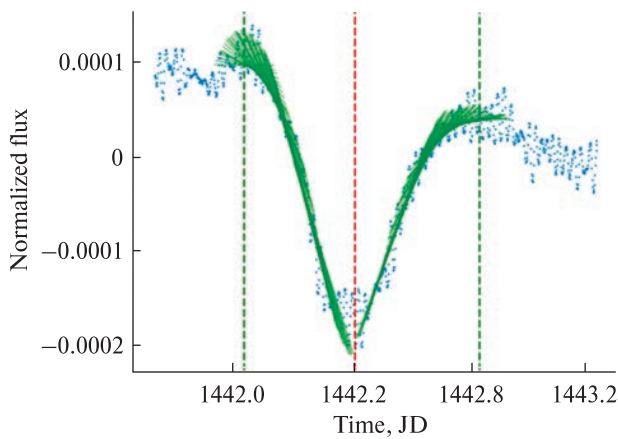

**Figure 11.** Linear approximations to find the ingress and egress points (vertical dashed green lines) moving away from the point of minimum (vertical dashed red line)

with durations between 0.3 and 3 days, corresponding roughly to half of a typical transit. For these selected minima, the positive segments on the right are

then analysed in the same manner:

$$w_{left\,max}(t) = \max\left\{ w \in W_{left} \mid \frac{1}{w}\sum_{i=t-w+1}^{t} 1_{\{y(i)>0\}} \le 0.05 \right\},$$

where $w_{left\,max}(t)$ is the maximum width of the window to the left of point $t$, where the conditions are satisfied; $W_{left}$ is the set of possible window widths {2, 1.95, 1.90, ..., 0.30}, decreasing by steps of 0.05; $y(i)$ represents the differenced data at point i; $1_{\{y(i)>0\}}$ is an indicator function that is 1 if $y(i)$ is positive and 0 otherwise; the sum $\sum_{i=t-w+1}^{t} 1_{\{y(i)>0\}}$ calculates the count of positive values within the window; $1/w$ normalizes the sum to the window width to find the proportion of positive values; and the inequality ≤0.05 ensures that the window contains no more than 5 % positive values, allowing for at least 95 % negative values within the window.

The left and right sides of a transit may differ in width, with the ascending part often broader as the comet's dust gradually disperses. Therefore, only cases where the right side is 1...4 times wider than the left are considered, which further reduces the number of candidates for manual inspection:

$$w_{right\,max}(t) = \max\left\{ w \in W_{right} \mid \frac{1}{w}\sum_{i=t}^{t+w} 1_{\{y(i)<0\}} \le 0.05 \right\},$$

where $w_{right\,max}(t)$ is the maximum width of the window to the right of point $t$, where the conditions are satisfied; $W_{right}$ is the set of possible window widths {4 * $W_{left}$, ..., $W_{left}$}, decreasing by steps of 0.05; $y(i)$ represents the differenced data at point i; $1_{\{y(i)<0\}}$, is an indicator function that is 1 if $y(i)$ is negative and 0 otherwise; the sum $\sum_{i=t}^{t+w} 1_{\{y(i)<0\}}$ calculates the count of negative values within the window; $1/w$ normalizes the sum to the window width to find the proportion of negative values; and the inequality ≤0.05 ensures that the window contains no more than 5 % negative values, allowing for at least 95 % positive values within the window.

To determine the ingress and egress of potential transits, we analysed the differenced light curves to find where the trend becomes negative on the left and positive on the right. Minor fluctuations near the transit edges can, however, cause false sign changes and lead to misinterpretation. A more reliable method applies a shifting linear regression to the time series, identifying ingress and egress points where the slope approaches zero.

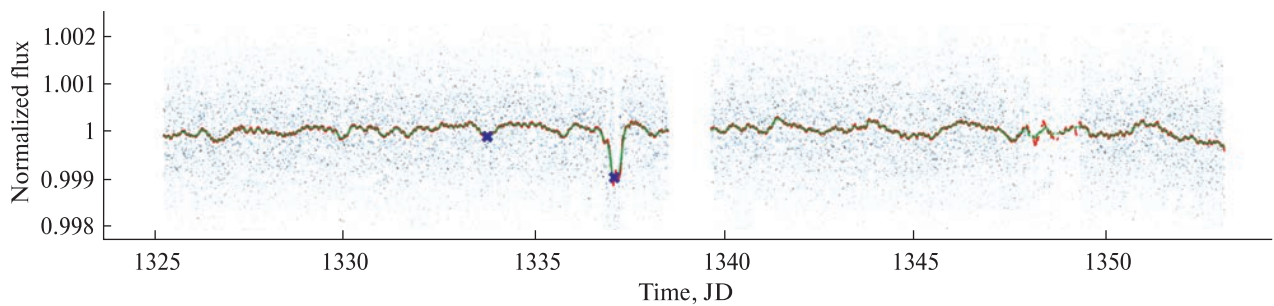

***Figure* 12**. Normalized flux for TIC 441042358 in TESS sector 1 with 2 marked exocomet candidates (X)

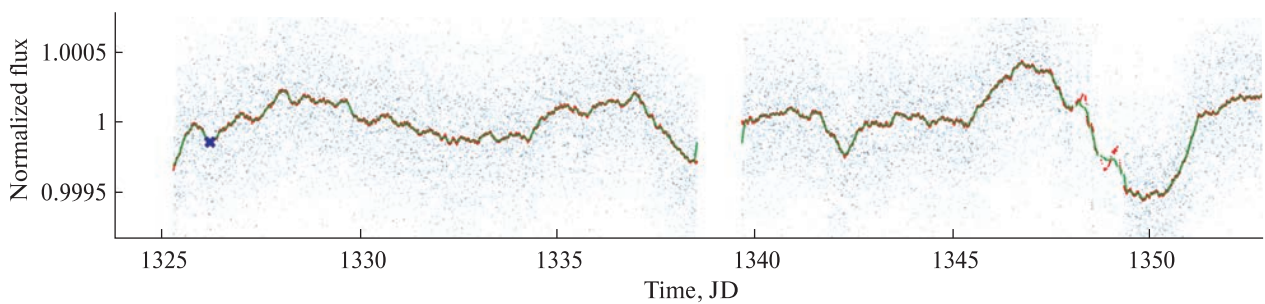

***Figure* 13**. Normalized flux for TIC 121490734 in TESS sector 1, with a marked presumably false-positive exocomet candidate (X)

We use the LinearRegression function from the scikit-learn library [21], moving from the flux minimum with a step $\Delta t$ and computing slopes within a centered window:

$$m = LinearRegression\left( flux\left[ t - \frac{W}{2}, t + \frac{W}{2} \right].slope \right)$$

if we are moving left from current $t$,
and $m_{left} < \theta$, then $t_{ingress} = t$,
if we are moving right from current $t$,
and $m_{right} < \theta$, then $t_{egress} = t$,

where $W$ is the time window size, $\Delta t$ is the step size used for moving through the time series, $\theta$ is the slope threshold for detecting ingress or egress, $t$ is the current time point being evaluated, $m$ is the slope of the linear regression line fitted to the data within the window centered at $t$, $m_{left}$ and $m_{right}$ represent the slope of the linear regression to the left and right of the time point $t$, respectively, $t_{ingress}$ and $t_{egress}$ are the required points of ingress and egress, respectively.

The values for the constants were set as follows: $W = 0.3$, $\Delta t = 0.005$, $\theta = 0.00001$. The process of finding the ingress and egress points with linear re-gression is presented in Figure 11, where the algorithm moves from the minimum point to the left and right separately, and stops when the slope of the approximated line is below the defined threshold.

## 5. RESULT OF THE STATISTICAL APPROACH

The described approach enables us to conduct a follow-up investigation into stars with presumed candidates for exocomets. Based on the two suspects identified by the machine learning approach (TIC 121490734 and TIC 441042358), we also ran the windowing algorithm to detect potential transit events and their minima precisely. Figure 12 shows the statistical analysis of TIC 441042358. This allowed us to confirm one potential asymmetrical transit event for TIC 441042358 at $T_0 = 1337.12$ BJD, which we found with visual inspection in paragraph 3 and Figure 6. At the same time, the algorithm detected a false-positive transit event at $T_0 = 1333.79$. This case shows that the statistical method is able to reproduce the same event that was found visually after applying the machine learning approach. It increases the confidence that the detected signal is real and not

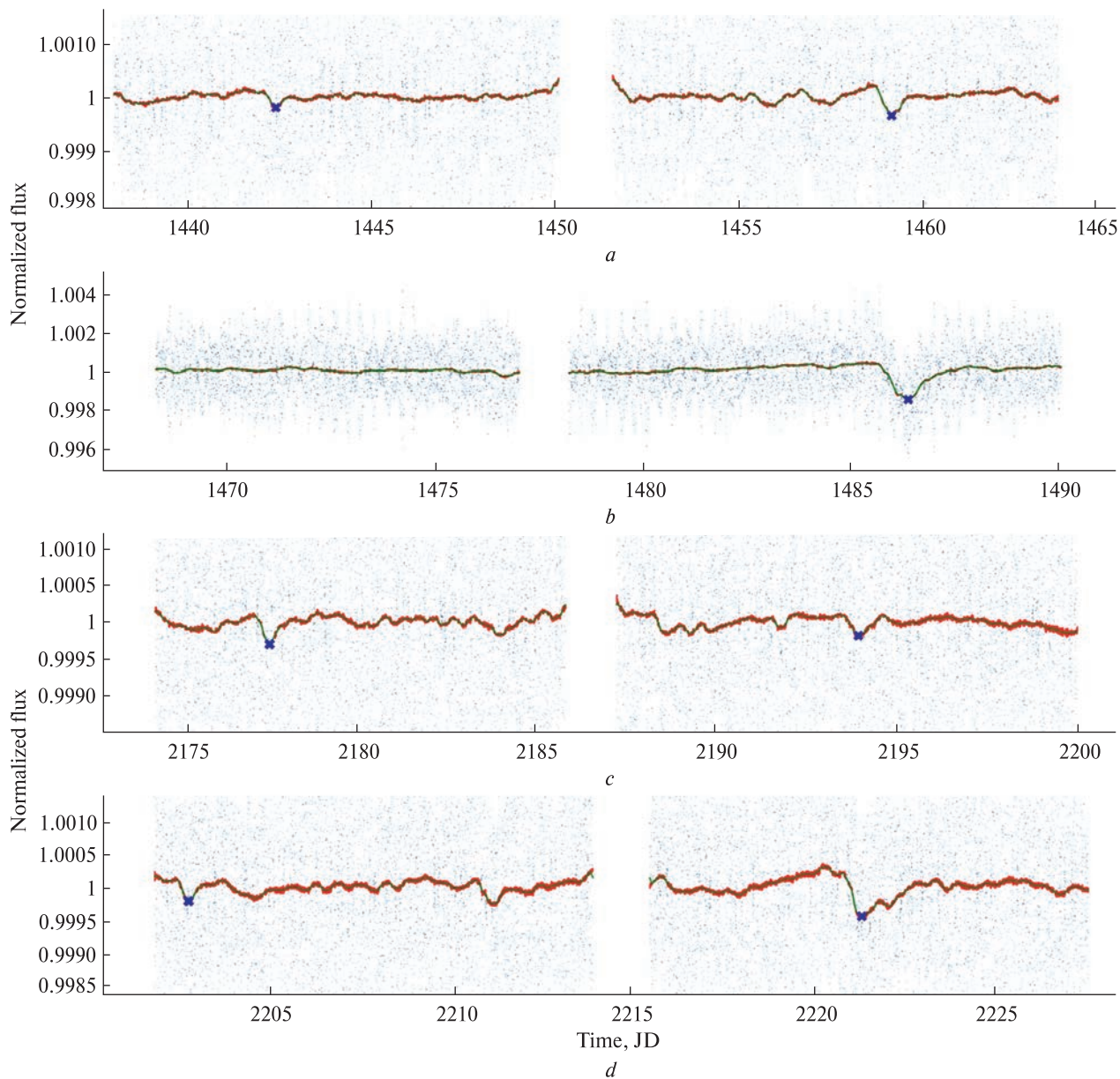

**Figure 14.** Normalized flux for TIC 270577175 (β Pictoris) in TESS sectors 5 (*a*), 6 (*b*), 32 (*c*), 33 (*d*) with marked exocomet candidates as blue X signs

random noise, and it also aids in better estimating the shape and duration of the asymmetry.

However, this approach hasn't detected any useful transit events in the flux of TIC 121490734 in sector 1, unlike our visual inspection (Figures 8, 9). Moreover, it additionally marked one false-positive event at $T_0 = 1326.220$ (Figure 13). Such a result shows that the method is quite sensitive to the chosen window size and to the background noise of the light curve. In general, it works well in filtering out random fluctuations, but it can miss weaker or shorter events that are still visible by eye. Therefore, combining the statistical check with visual analysis remains the best way to confirm faint or shorter asymmetric transits.

To further test the windowing approach, we also applied it to the β Pictoris light curves in different sectors, where exocomet candidates are known to exist [12, 13, 19, 30]. The previous research of the star's light curves identified 30 exocomet transits with transit depth between 0.01 % to 0.04 % of the star's flux [13]. From this sample the deeper ones are following: the 2 events in sector 5 at $T_0 = 1442.37 \pm 0.02$ BJD, and $1459.16 \pm 0.02$ BJD (Fig. 14, *a*); 1 event in sector 6 at $1486.40 \pm 0.01$ BJD (Fig. 14, *b*); 2 events in sector 32 at $T_0 = 2177.45 \pm 0.03$ BJD and $2193.98 \pm 0.02$ BJD (Fig. 14, *c*); and 3 events in sector 33 at $T_0 = 2202.71 \pm 0.02$ BJD, $2211.11 \pm 0.01$, and $2221.4 \pm 0.1$ BJD (Fig. 14, *d*), which are deeper than 0.03 % of the star's flux. The proposed statistical approach managed to capture almost all events from the papers mentioned above in sectors 5, 6, 32, and 33. Among the three events from sector 33, it didn't capture only the middle one at $T_0 = 2211.11 \pm 0.01$ BJD. Consequently, according to [13]'s report, all events deeper than $0.035 \pm 0.003$ % of the star flux were identified except two ones at 1476.34 and 2211.11 BJD.

The fact that the algorithm reproduced almost all known events demonstrates that the method works in different sectors and under various noise conditions. These results also show that it could be applied to other bright TESS targets or to future missions to search for similar short, asymmetric events. In general, the simple statistical approach supports the machine learning model's detections, provides quantitative verification of the results, and helps separate actual cometary transits from noise or instrumental variations. Together with the visual inspection, it provides a more comprehensive way to search for new exocomet candidates.

## 6. CONCLUSION

The ML Random Forest method allows potentially to identify asymmetric minima in the light curves due to the transits of comet-like bodies with an accuracy of 97 % if the minimum depth is larger than 0.02 % of the star's flux and for light curves with a noise level corresponding to CDPP < 150. Since the detection of any signal becomes more difficult with increasing noise level, our result is the first step towards the ability to detect very shallow comet transits in the light curves of poorer quality.

The ML approach allowed us to find only two suspicious events in sector 1, and both of them have depth less than 0.02 % of the star's flux. It should be noted that one candidate for an exocomet transit was found by [17] in this sector in the TIC 280832588 light curve, but this star was not in our database because about 32 % of cadences were flagged, pointing out some instrumental artifacts.

In order to reinforce the ML approach, the statistical method was used, which identified a suspicious event in the TIC 441042358 light curve, but missed the event in the TIC 121490734 light curve. Additionally, the statistical method found almost all transits previously identified in the light curve of β Pictoris in 5, 6, 32, 33, which are deeper than 0.03 % of the star's flux. Failing to identify all exocomet candidates found and marking some false-positive transits signifies the importance of properly setting the constant parameters, which we defined for the system using the statistical method. Further fine tuning these parameters to find the best values, or even introducing some dynamic mechanism for their calculation, should improve the quality of the provided results of the execution.

An additional challenge is the intrinsic variability of stars and the presence of star spots, which make the identification of shallow exocomet transits more complicated. Future improvements to the algorithm will focus on separating and extracting the stellar harmonic oscillations before applying machine learning techniques, as well as using a wider set of features for the classifier and more powerful deep neural networks. Nevertheless, two potential candidates have been identified that require further observations to confirm these events as potential exocomet candidates.

The combined use of the ML and statistical approach demonstrates that it can be applied for detecting exocometary dips in TESS data. The cross-validation of both methods helps to reduce the number of false positives and increases confidence in identifying truly asymmetric profiles.

*Д. В. Добричева*[1], старш. наук. співроб, канд. фіз.-мат. наук , старш. дослід.

*І. В. Кулик*[1], старш. наук. співроб., канд. фіз.-мат. наук, старш. дослід.

*Д. Р. Каракуц*[1,2], мол. наук. співроб., аспірант

*М. Ю. Василенко*[1,2], мол. наук. співроб., аспірант

*Я. В. Павленко*[1], д-р фіз.-мат. наук

*О. С. Шубіна*[1,3], старш. наук. співроб., канд. фіз.-мат. наук

*І. В. Лук'яник*[4], зав. сектору, канд. фіз.-мат. наук, старш. дослід.

[1] Головна астрономічна обсерваторія Національної академії наук України

вул. Академіка Заболотного 27, Київ, Україна, 03143

[2] Інститут фізики Національної академії наук України

проспект Науки 46, Київ, Україна, 03028

[3] Астрономічний Інститут Словацької академії наук

Високі Татри, Татранська Ломниця, Словацька Республіка, 059 60,

[4] Астрономічна обсерваторія Київського національного університету імені Тараса Шевченка

вул. Обсерваторна 3, Київ, Україна, 04053

ВІЗУАЛЬНА ІНСПЕКЦІЯ ПОТЕНЦІЙНИХ ТРАНЗИТІВ ЕКЗОКОМЕТ,
ВИЯВЛЕНИХ ЗА ДОПОМОГОЮ МАШИННОГО НАВЧАННЯ ТА СТАТИСТИЧНИХ МЕТОДІВ

Розглядаються кілька методів виявлення можливих транзитів екзокомет у кривих блиску зір з бази даних орбітального телескопа TESS (The Transiting Exoplanet Survey Satellite). Перший метод заснований на алгоритмі випадкового лісу, одному із методів машинного навчання, було застосовано у попередній роботі авторів. Для тренування методу використовувалась вибірка змодельованих асиметричних мінімумів у кривих блиску зір, побудованих на основі розрахунків проходження модельного зображення комети по диску зорі. Після тренування метод було застосовано до реальних даних із першого сектора спостережень бази даних TESS. Даний підхід дозволив нам знайти 32 кандидати зі слабкими та непеіодичними зменшеннями яскравості зір, які можуть відповідати проходженням кометоподібних тіл по диску зорі. Метою даної роботи є аналіз даних кандидатів за допомогою візуальної інспекції, щоб переконатися, що виявлені особливості не були спричинені інструментальними ефектами.

Другий підхід, який ми пропонуємо для виявлення можливих транзитів екзокомет — незалежний статистичний метод для перевірки результатів та пошуку асиметричних мінімумів безпосередньо у кривих блиску. Цей підхід було застосовано до кривих блиску зорі β Живописця у секторах 5, 6, 32 та 33 спостережень TESS. Алгоритм відтворив майже всі раніше відомі події, глибина яких була більшою за 0.03 % потоку зорі. Це демонструє ефективність методу для пошуку неглибоких та нерегулярних змін потоку зір у кривих блиску з різних секторів і, відповідно, з різним рівнем зашумленості.

Поєднання методів машинного навчання, візуального огляду та незалежного статистичного аналізу кривих блиску зір полегшує ідентифікацію слабких та короткотривалих подій, які проявляють себе як асиметричні транзити у фотометричних даних. Хоча кількість підтверджених транзитів екзокомет все ще невелика, зрослий масив спостережень вказує на їхню ймовірну наявність у багатьох молодих планетних системах.

***Ключові слова:*** комети, планетні системи, малі планети; затемнення, транзити, планети та супутники; методи машинного навчання, статистичні методи, візуальна інспекція.